\RequirePackage{fix-cm}
\documentclass[twocolumn]{svjour3}          
\smartqed  
\usepackage{graphicx}
\begin{document}

\title{Electronic and magnetic properties of the BaTiO$_{3}$/LaMnO$_{3}$ interface: a DFT study}
\thanks{The reported study was funded by the Russian Scientific Foundation according to research project No. 21-12-00179.
}


\author{Irina Piyanzina         \and
       Rinat Mamin 
}


\institute{I. Piyanzina and R.Mamin \at
               Zavoisky Physical-Technical Institute, FIC KazanSC of
	RAS, 420029 Kazan, Russia \\
              Tel.: +79196931320\\
              \email{i.piyanzina@gmail.com}           
           \and
}

\date{Received: date / Accepted: date}

\maketitle

\begin{abstract}
By means of \textit{ab initio} calculations withing the density functional theory (DFT)  electronic and magnetic properties of BaTiO$_{3}$/LaMnO$_{3}$ interface were investigated. An impact of ferroelectric overlayer thickness onto the interface properties was analysed through the spin-polarized density of states.
\keywords{ferroelectric interface \and magnetoelectric coupling \and DFT}
\end{abstract}

\section{Introduction}
\label{intro}
The LaAlO$_{3}$/SrTiO$_{3}$ (LAO/STO) heterostructure has been widely investigated during last 20 years after the two-dimensional gas (2DEG) observation at the interface of that system~\cite{ohtomo}. It was also found the conducting state coexists with magnetic state.  The arising magnetic order in the LAO/STO system is a matter of intensive discussion~\cite{pavlenko,michaeli}. From \textit{ab initio} calculations it was revealed that the bare heterostructure is non-magnetic, and magnetic ordering was related to defects formation, in particular, to oxygen vacancies at the interface and/or the surface~\cite{pavlenko,piyanzina}. The arising magnetization is weak and in according to Ref.~\cite{piyanzina} the maximal magnetic moment per Ti atom equals to 0.232~$\mu_{B}$. 

Besides, below 300\,mK LAO/STO system passes into the superconducting state~\cite{reyren}. Recently, in one of our previous research by means of \textit{ab initio} it was demonstrated that the usage of the high-temperature superconductor (PCHTSC), La$_{2}$CuO$_{4}$ (LCO), as a substrate and ferroelectric oxide BaTiO$_{3}$ in the heterostructure as an overlayer may result in conducting state located mostly at the interfacial LCO layer. Lately, a high-temperature quasi-two-dimensional superconducting state has been observed in the Ba$_{0.8}$Sr$_{0.2}$TiO$_{3}$/La$_{2}$CuO$_{4}$ heterostructure with $T$$_{c}$=30\,K~\cite{pavlov}. This $T$$_{c}$ is 100 times larger than Tc in LAO/STO~\cite{reyren}.

The experimental observation of conducting state was also realized in the Ba$_{0.8}$Sr$_{0.2}$TiO$_3$/
LaMnO$_3$ system~\cite{pavlovJETP}.
Such a heterostructute is composed of a  ferromagnet which can serve as a source of magnetic state and a antiferroelectric  which can be used as a tool for electron doping~\cite{weng,kabanov}.
Hence, in the LaMnO$_{3}$/BaTiO$_{3}$ (LMO/BTO) heterointerface it is expected to have the interfacial magnetism along with conducting state.
It presented paper we  investigate the possibilities of the spin-polarized 2DEG appearing in the LMO/BTO heterostructure. The essential issue  is to understand the impact of the ferroelectric BTO slab thickness on the interface electronic and magnetic states. In order to do so we investigate the BTO/LMO heterostructure with varying number of BTO overlayers and analyse the density of state spectrum. 

\section{Calculation details}
\label{method}
The {\it ab initio} calculations were based on density functional  theory (DFT)~\cite{hohenberg1964,kohn1965}. Exchange and correlation effects were accounted for by the generalized gradient approximation (GGA) as parametrized by Perdew, Burke, and Ernzerhof (PBE)~\cite{perdew1996}. The Kohn-Sham equations were solved with projector-augmented-wave (PAW) potentials and wave functions~\cite{bloechl1994paw} as implemented in the Vienna Ab-Initio Simulation Package (VASP)~\cite{kresse1996a,kresse1996b,kresse1999}, which is part of the MedeA\textsuperscript{\textregistered} software of Materials Design~\cite{medea}. Specifically, we used a plane-wave cutoff of 400\,eV. The force tolerance was 0.05\,eV/\AA\ and the energy tolerance for the self-consistency loop was $ 10^{-5} $\,eV. The Brillouin zones were sampled using Monkhorst-Pack grids~\cite{monkhorst1976} including $ 5 \times 5 \times 1 $ $ {\bf k} $-points.
A set of calculations was carried out with a \textit{+U} correction applied to Mn 3\textit{d}, Ti 3\textit{d} and La 4\textit{f} states~\cite{piyanzina2017}. A simplified Dudarev approach was used~\cite{dudarev1998}: the \textit{U}  values of 4\,eV  for Mn,  2\,eV for Ti  and 8\,eV for La states were applied.

The heterostructures were modelled by a central region of $ {\rm LaMnO_3} $ comprising $ 2 \frac{1}{2} $ unit cells with $ {\rm LaO} $ termination on both sides and varying number of  $ {\rm BaTiO_3} $ overlayers  with $ {\rm TiO_{2}} $ termination towards the central slab and $ {\rm BaO} $ surface termination also on both sides.
In order to avoid interaction of the surfaces and slabs with their periodic images, $\approx$20~\AA-wide vacuum region was added.

\section{Bulk components}
\label{bulk}

First, the parent materials of the heterointerface have been checked separately in the bulk geometry in order to ensure the reproducibility of the results obtained by the method and computational parameters used in the present research.

\begin{figure}[h]
 \includegraphics[width=0.49\textwidth]{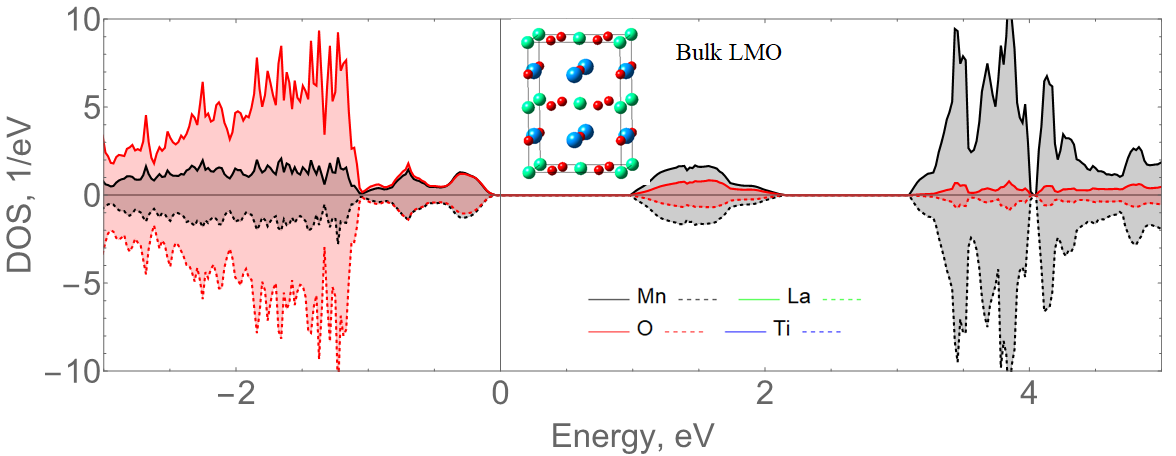}
  \caption{Density of states spectra for the A-AFM bulk LaMnO$_{3}$ along with the corresponding unit cell structure.}
\centering
\label{img:Figure1}
\end{figure}

Starting from the experimental structure of bulk LaMnO$_{3}$, the lattice constants and atomic positions were fully relaxed.
%
%
To realize the spin-dependent switching effect, which was mentioned in the introduction, it was suggested to replace FM material by AFM~\cite{weng}.  The spin-polarized density of states (DOS) plot for the A-AFM LMO in the bulk configuration in presented in Fig.~\ref{img:Figure1}. The bulk LMO is a semiconductor, with an O~2\textit{p} dominated valence band, with Mn 3\textit{d} contribution. The calculated band gap and magnetic moment of the Mn atom equal 1.349\,eV and 3.832\,$\mu_{B}$, respectively. Those values agree well with experimental ones of 1.7\,eV and $3.7\pm0.1$\,$\mu_{B}$, respectively~\cite{saitoh}. Based on this comparison we concluded that the chosen value of the \textit{U} parameter can yield relatively correctly both the energy gap and the magnetization. Besides, the calculated cell parameters, shown in Table~\ref{table2}, were found to be close to experimental values~\cite{sawada} and also to the previous \textit{ab initio} research, for example Ref.~\cite{ciucivara2008}. 
\begin{table}[h]
\centering
\caption{Calculated lattice constants in \AA\ of bulk LaMnO$_{3}$, bulk  BaTiO$_{3}$ and a rotated by 45$^{\circ}$ along \textit{z}-axis BaTiO$_{3}$ unit cell in order to merge with  the LaMnO$_{3}$ unit cell. The last row lists the BaTiO$_{3}$/LaMnO$_{3}$ supercell lattice constants. Experimental data are presented as well for comparison.}
\label{table2}
\begin{tabular*}{0.35\textwidth}{c | c c c  } 
\hline 
 & a & b & c  \\ 
\hline 
LMO & ~5.709~ & ~5.675~ & ~8.018~   \\ 
Expt.~\cite{sawada} & 5.742 & 5.532 & 7.669 \\ 
\hline 
BTO & 3.986 & 3.986 & 4.014   \\ 
Expt.~\cite{crystal} & 3.992 & 3.992 & 4.036  \\ 
BTO ($\times \sqrt{2}$) & 5.637 & 5.637 & 4.014 \\  
\hline 
Supercell & 5.709 & 5.675 & 50  \\
\hline 
\end{tabular*}
\end{table}


BaTiO$_{3}$ is one of the most well known ferroelectric, which has a ferroelectric polarization in a tetragonal system with moderate polarization of 26\,$\mu$C/cm$^{2}$~\cite{subarao}.  
%
\begin{table}[h]
\centering
\caption{Calculated energy ($\textit{E}$) per unit cell,  c/a ratio, the band gaps ($\varepsilon$), displacement of Ti atoms with respect to the O planes ($\bigtriangleup$) and polarization $\textbf{P}$ in $\mu$C/cm$^{2}$  of the cubic, tetragonal and orthorombic phases of bulk  BaTiO$_{3}$. Experimental values are given for the tetragonal phase.}
\label{table3}
\begin{tabular*}{0.49\textwidth}{c | c | c | c | c } 
\hline 
Phase &  c/a & $\varepsilon$,\,eV & $\bigtriangleup$,\,\AA & $\textbf{P}$ \\ 
\hline
Cubic &  1 & 2.169 & 0 & 0 \\ 
\hline
Tetragonal &  1.007  & 2.249  & 0.13 & 31   \\ 
Expt.  & 1.010\,\cite{shirane} & 3.27\,\cite{wemple}& 0.15\,\cite{shirane} & 26\,\cite{subarao} \\
\hline
Orthorhombic &  1.428 & 2.259 & 0.09 &   \\  
\hline 
\end{tabular*}
\end{table}
\begin{figure}[ht]
 \includegraphics [width=0.49\textwidth]{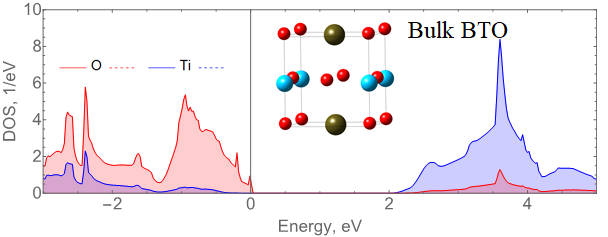}
  \caption{Density of states spectra for the bulk  BaTiO$_{3}$ along with the corresponding unit cell structure.}
\centering
\label{img:Figure2}
\end{figure}
Calculated energies per unit cells, band gaps, oxygen displacements and polarization for the cubic, tetragonal and orthorhombic phases are presented in Table~\ref{table3}. All calculated values agree well with experimental data. The experimental band gap is higher than the computed one, but the difference is reasonable for the DFT. For the purpose of the present work we are interested in the phases with spontaneous polarization, hence, we will focus on the tetragonal BTO structure. The calculated lattice parameters along with experimental ones are listed in Table~\ref{table2}.
The calculated density of state spectra for the bulk BTO is given in Fig.~\ref{img:Figure2}. 

\subsection{LaMnO$_{3}$/BaTiO$_{3}$ heterostructure}
\label{hetero}
\begin{figure}[ht]
 \includegraphics[width=0.5\textwidth]  {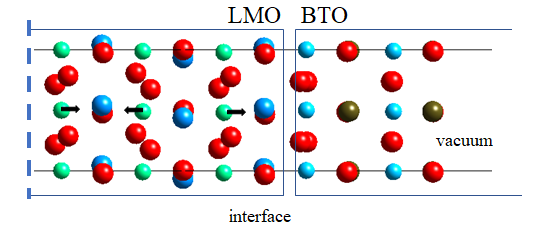}
  \caption{ The fully optimized half-right structure of LMO/2BTO heterointerface. The supercell is displayed without the vacuum region, which was ~20\AA. Arrows indicate magnetic moments directions in the LMO slab. b) Corresponding atom-resolved density of states for the heterostructure with two BTO overlayer and c) with three BTO overlayers.}
\centering
\label{img:cell}
\end{figure}

In order to merge  BTO with LMO so that the polarization is parallel to the easy axis of antiferromagnet, the BTO unit cell has to be rotated by  45$^{\circ}$ along \textit{z}-axis. As listed in Table\,\ref{table2}  $a_{BTO}\times\sqrt{2}$ are very close to the $a_{LMO}$ and $b_{LMO}$ cell parameters. The resulted supercell with two BTO overlayers is presented in Fig.~\ref{img:cell}\,a, where the right half of the unit cell is presented without full vacuum region. The structure was fully optimized.

\begin{figure}[ht]
 \includegraphics[width=0.5\textwidth]  {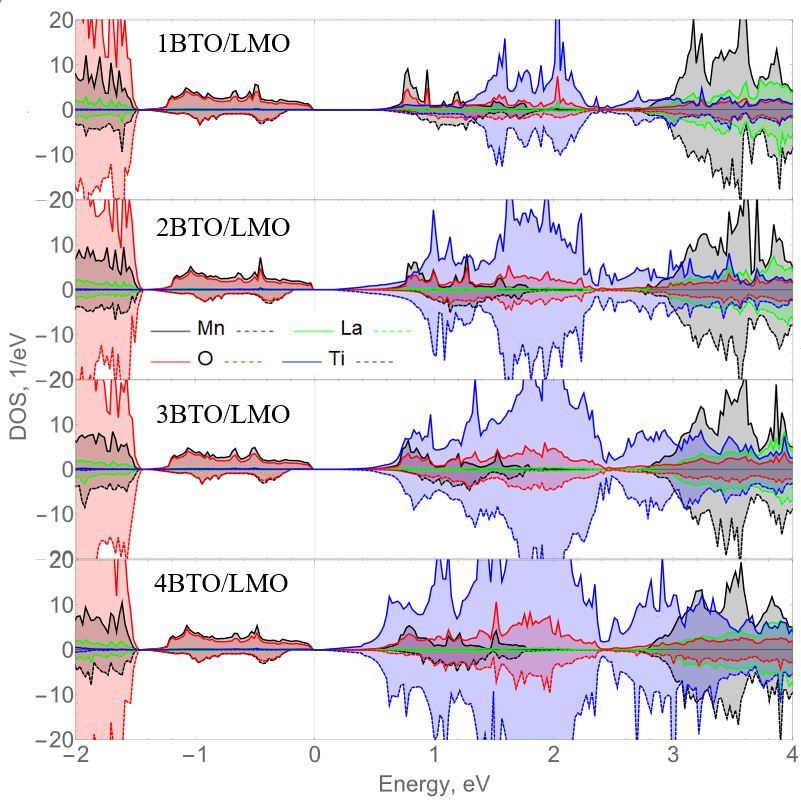}
  \caption{Atom-resolved density of states spectra for the heterostructure with varying number (1-4) BTO overlayers.}
\centering
\label{img:dos}
\end{figure}

\begin{figure}[ht]
 \includegraphics [width=0.49\textwidth]{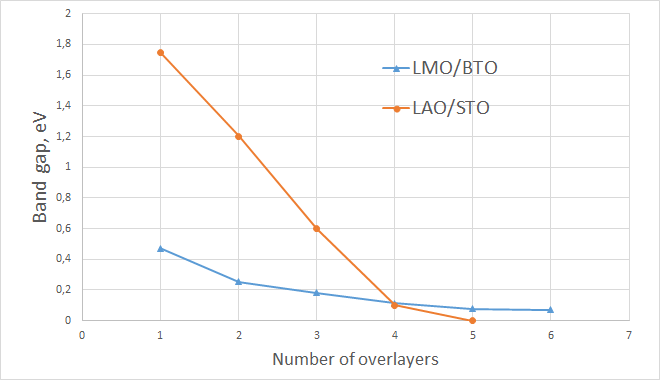}
  \caption{Calculated band gap versus the number of BTO  overlayers in LMO/BTO heterostructure along with analogous dependence for LAO/STO systems for comparison.}
\centering
\label{img:Figure5}
\end{figure}

Performed structural optimization resulted in the insignificant shift of Ti atoms out of oxygen planes ($\Delta$\,z$_{Ti-O}$). For the heterostructure with two BTO overlayers  $\Delta$\,z$_{Ti-O}$ distances within the interfacial layer equal to -0.046\,\AA\ and 0.247\,\AA, within the surface layer shifts are less significant and equal to 0.049\,\AA\ and 0.101\,\AA , respectively.  Except for the one, all the movements of Ti are towards the surface, which leads to the total polarization predominantly towards the surface. Such a structural reconstruction leads to the electronic rearrangement as reflected in the DOS spectra shown in Fig.~\ref{img:dos}\,a\,--\,d for varying number of BTO layers. With increasing the number of BTO overlayers the band gap decreases from 0.47\,eV for heterostructure with one BTO overlayer to 0.113\,eV with four. Such a decay of the band gap with the number of overlayers differs from LAO/STO linear dependence as shown in Fig.~\ref{img:Figure5}. Indeed, the band gap decrease with increase of LAO overlayers in LAO/STO associated with increase of the field directed towards the surface. As soon as this field overcomes the field arising due do the structural relaxation, the LAO/STO system becomes conductor. That happens above 4 LAO overlayers in according to our previous \textit{ab initio} research~\cite{piyanzina}. In contract, in the LMO/BTO heterostructure without polar overlayers the field towards the surface originates from the ferroelectric polarization. Since the slope in Fig.~\ref{img:Figure5} for LMO/BTO in smaller it is obvious that the ferroelectric field is weaker and conductivity occurs with more overlayers. Note, that in experiments~\cite{pavlovJETP} the thickness of the ferroelectric film was about 350\,nm, and based on the asymptotics of our calculations it is expected that the band gap vanishes completely at such a thickness. 

Finally, the LMO/BTO system possesses relatively high total magnetization due to the odd number of MnO layers with magnetic moment per Mn equal to 3.8~$\mu_{B}$. Ferromagnetic magnetization should increase with increasing the electron doping.

\section{Conclusions}
In the present paper by means of DFT+$U$ calculation the electronic and magnetic properties of bulk LaMnO$_3$  and  BaTiO$_3$, as well as  LMO/BTO heterostructure have been demonstrated. Within the chosen approach and computational parameters the bulk components of the heterostructure were confirmed to be insulators. In the heterostructure geometry the  decrease of the band gap with increasing the number of BTO overlayers was demonstrated. It was found that the curve tends to zero, but the system remains semiconductor up to six BTO overlayers. It means that the conducting state arises with more ferroelectric overlayers, that is consistent with experiment from Ref.~\cite{pavlovJETP} where the thickness of ferroelectric film was much higher. It was shown that the LMO/BTO systen possesses relatively high total magnetization, which is expected to increase with increasing the electron doping.

\begin{acknowledgements}
The reported study was funded by the Russian Scientific Foundation according to research project No. 21-12-00179.
\end{acknowledgements}

%
\section*{Conflict of interest}
The authors declare that they have no conflict of interest.



\end{document}